\documentclass[
 aps,
 pre,
 reprint,
 superscriptaddress,
 nofootinbib,
 amsmath,
 amssymb,
 floatfix
]{revtex4-2}

\usepackage[utf8]{inputenc}
\usepackage{graphicx}
\usepackage{bm}
\usepackage{bbm}
\usepackage{braket}
\usepackage{booktabs}
\usepackage{dcolumn}
\usepackage{placeins}
\usepackage{capt-of}

\graphicspath{{images/}{}}

\DeclareMathOperator{\Tr}{Tr}
\newcommand{\dd}{\mathrm{d}}
\newcommand{\one}{\mathbbm{1}}
\newcommand{\rvec}{\mathbf{r}}
\newcommand{\xvec}{\mathbf{x}}
\newcommand{\Rvec}{\mathbf{R}}
\newcommand{\RDM}{\rho^{(1)}}
\newcommand{\FRDMF}{F_{\mathrm{RDMF}}}
\newcommand{\FDF}{F_{\mathrm{DF}}}

\begin{document}

\title{Reducing the Complexity of Density-Matrix Functionals in a Real-Space-Decomposed DF+RDMF Scheme with the Adaptive Cluster Approximation}

\author{Konstantin Tamoev}
\affiliation{Dynamics of Condensed Matter, Department of Chemistry, Paderborn University, Warburger Str. 100, 33098 Paderborn, Germany}

\author{Robert Schade}
\affiliation{Paderborn Center for Parallel Computing (PC$^2$), Paderborn University, Warburger Str. 100, 33098 Paderborn, Germany}

\author{Thomas D. K\"uhne}
\email{tkuehne@cp2k.org}
\affiliation{Center for Advanced Systems Understanding (CASUS), Helmholtz Zentrum Dresden-Rossendorf, Germany}
\affiliation{Institute of Artificial Intelligence, Technische Universit\"at Dresden, Germany}

\date{\today}

\begin{abstract}
Reduced density-matrix functional theory (RDMFT) provides a variational route to electronic correlations beyond conventional density-functional approximations, but explicit evaluations of density-matrix functionals still scale exponentially with the number of active one-particle states.  We formulate and assess a real-space-decomposed density-functional plus reduced-density-matrix-functional (DF+RDMF) scheme in which the Coulomb interaction is partitioned locally in real space and the RDMF correction is evaluated only for the strongly correlated part of the interaction.  The resulting local density-matrix functionals are further compressed using the adaptive cluster approximation (ACA), which performs a unitary rotation of the bath subspace before truncation and therefore preserves the local interaction while reducing the number of explicitly correlated bath states.  As a molecular test case, we consider the bending potential of carbon suboxide, C$_3$O$_2$.  While semilocal PBE favors a linear molecule, the DF+RDMF/ACA correction stabilizes a bent configuration in qualitative agreement with the quasilinear behavior inferred from spectroscopy.  The approach provides a systematic embedding hierarchy for combining density functionals with explicitly correlated density-matrix corrections in extended or spatially inhomogeneous systems.
\end{abstract}

\maketitle

\section{Introduction}

Electronic-structure methods must balance accuracy, transferability, and computational cost.  Wave-function-based quantum-chemical methods provide systematic access to electronic correlation, but their scaling with system size severely limits routine applications to extended systems \cite{Pople1999}.  Density-functional theory (DFT), by contrast, is efficient and broadly applicable \cite{Kohn1999}; however, the use of the electron density as its basic variable makes the construction of robust approximations for strongly correlated situations difficult \cite{Pernal2015}.

Reduced density-matrix functional theory (RDMFT) offers an appealing alternative because the one-particle reduced density matrix (1RDM) is nonlocal and contains direct information about fractional natural-orbital occupations \cite{Gilbert1975}.  In particular, the kinetic energy is known explicitly as a functional of the 1RDM, while the unknown density-matrix functional contains only interaction and entropic contributions.  This structure has motivated approximate natural-orbital functionals, from the M{\"u}ller ansatz to the Goedecker--Umrigar correction, and makes RDMFT a natural framework for describing local strong correlations \cite{Muller1984,GoedeckerUmrigar1998}.  The practical obstacle is that accurate density-matrix functionals are difficult to evaluate: exact or near-exact evaluations require a constrained search over many-particle states and therefore scale exponentially with the number of one-particle states included in the correlated subspace.  This bottleneck has also motivated hybrid quantum-classical RDMFT algorithms, in which effective subsystem functionals are evaluated on quantum hardware while the decomposition and optimization remain classical \cite{Schade2022ParallelQC}.

Subsequent developments have broadened this landscape substantially.  Physically motivated corrections and Piris-type natural-orbital functionals improved molecular energetics, while benchmark and density-matrix-power studies clarified the practical accuracy of RDMFT for finite systems and the homogeneous electron gas \cite{GritsenkoPernalBaerends2005,Piris2021,LathiotakisMarques2008,LathiotakisSharma2009}.  The formalism has also been extended to linear-response and time-dependent RDMFT, and to correlated solids where power-functionals and related approaches have been used for Mott-insulator physics \cite{PernalGiesbertz2007,GiesbertzGritsenkoBaerends2010,Sharma2008,Sharma2013}.

One route around this obstacle is to combine density functionals and reduced-density-matrix functionals.  In such DF+RDMF schemes, the weakly correlated and long-range part of the interaction is described by a density functional, whereas a density-matrix correction is retained for the locally important interaction channels \cite{Bloechl2011,schade2019new}.  The key question is then how to define the local interaction and how to evaluate the associated density-matrix functional at affordable computational cost.

Here we refine this idea in two steps.  First, we use a real-space decomposition of the Coulomb interaction, which avoids the double-counting ambiguities associated with purely orbital-based interaction partitions.  Second, we reduce the remaining correlated problem by the adaptive cluster approximation (ACA) \cite{aca_schade_bloechl_2018,Schade2022ParallelQC}.  The ACA performs a unitary transformation of the noninteracting bath states such that truncating distant bath levels introduces a controlled and systematically improvable approximation.  We apply this framework to the bending potential of C$_3$O$_2$, a compact test case in which local correlation around the central carbon atom can be isolated and contrasted with a semilocal PBE density-functional baseline in CP-PAW \cite{PerdewBurkeErnzerhof1996,Bloechl2026CPPAW}.

The manuscript is organized as follows.  Section~\ref{sec:rdmft} summarizes the RDMFT formalism and the DF+RDMF decomposition.  Section~\ref{sec:realspace} introduces the real-space interaction partition.  Section~\ref{sec:aca} describes the ACA reduction of the local density-matrix functional.  Section~\ref{sec:results} presents the C$_3$O$_2$ application, and Sec.~\ref{sec:conclusions} gives the conclusions.

\section{Reduced Density-Matrix Functional Theory}
\label{sec:rdmft}

RDMFT can be viewed as a variational electronic-structure theory in which the basic variable is promoted from the density $n(\rvec)$ to the one-particle reduced density matrix.  In coordinate-spin notation, $\xvec=(\rvec,\sigma)$, the 1RDM admits the natural-orbital representation
\begin{equation}
\RDM(\xvec,\xvec')
=\sum_i n_i\varphi_i(\xvec)\varphi_i^*(\xvec'),
\label{eq:natural_orbitals}
\end{equation}
where the natural occupation numbers obey $0\leq n_i\leq 1$ and $\sum_i n_i=N$ for an ensemble $N$-representable state \cite{Coleman1963}.  The electron density is recovered from the diagonal,
\begin{equation}
n(\rvec)=\sum_\sigma \RDM(\rvec\sigma,\rvec\sigma),
\end{equation}
but the off-diagonal structure of $\RDM$ and the fractional occupations provide information about delocalization, exchange, and static correlation that is absent from the density alone \cite{Gilbert1975,GoedeckerUmrigar1998,Pernal2015}.

The representability of reduced density matrices is a central constraint in any variational RDM theory.  For the 1RDM, Coleman's ensemble conditions are simple, while pure-state representability imposes additional generalized Pauli constraints on the natural occupation numbers \cite{Schilling2013Pinning}.  Finite-basis and finite-temperature formulations have further clarified the associated $v$-representability and differentiability issues \cite{GiesbertzRuggenthaler2019}.  For higher-order RDMs the representability problem is substantially more involved and remains a major organizing principle for wave-function-free electronic-structure methods \cite{Mazziotti2012NRepresentability}.

This choice of basic variable changes where approximations enter.  The one-particle contribution, including the kinetic energy, is explicit once $\RDM$ is known, whereas all interaction and entropic many-body contributions are collected in a universal functional of $\RDM$.  Exact evaluation of this functional for a nontrivial interaction is still a many-particle problem.  For $N$ electrons in $N_\chi$ spin orbitals, the fixed-particle-number Hilbert space has dimension $\binom{N_\chi}{N}$, while a grand-canonical or finite-temperature search involves the full Fock space of dimension $2^{N_\chi}$.  This motivates the DF+RDMF strategy used below: a density functional supplies the smooth, weakly correlated background, while an explicit reduced-density-matrix functional is retained only for the local interaction channels that require a correlated treatment.

\subsection{Variational formulation}

Let $h$ denote the one-particle Hamiltonian and $W$ the electron-electron interaction.  At inverse temperature $\beta=(k_{\mathrm{B}}T)^{-1}$ and chemical potential $\mu$, the grand potential can be written as the Legendre-Fenchel transform of the universal density-matrix functional $F_\beta^W[\RDM]$,
\begin{equation}
\Omega_{\beta,\mu}[h]
= \min_{0\leq \RDM \leq \one}
\left\{
\Tr\left[\RDM(h-\mu\one)\right]
+F_\beta^W[\RDM]
\right\}.
\label{eq:grand_potential}
\end{equation}
At fixed particle number $N$, the Helmholtz free energy is
\begin{equation}
H_{\beta,N}[h]
= \min_{\substack{0\leq \RDM \leq \one\\ \Tr[\RDM]=N}}
\left\{
\Tr[\RDM h]+F_\beta^W[\RDM]
\right\}.
\label{eq:helmholtz}
\end{equation}
In the zero-temperature limit this reduces to the ground-state variational principle
\begin{equation}
E_N[h]
= \min_{\substack{0\leq \RDM \leq \one\\ \Tr[\RDM]=N}}
\left\{
\Tr[\RDM h]+F^W[\RDM]
\right\}.
\label{eq:ground_state_energy}
\end{equation}
Conversely, at $\mu=0$ the density-matrix functional follows from
\begin{equation}
F_\beta^W[\RDM]
= \max_h
\left[
\Omega_{\beta,0}[h]-\Tr[\RDM h]
\right],
\label{eq:legendre}
\end{equation}
which identifies $h$ and $\RDM$ as conjugate variables,
\begin{align}
\frac{\partial \Omega_{\beta,\mu}[h]}{\partial h_{\alpha\beta}}
&= \RDM_{\beta\alpha},\\
\frac{\partial F_\beta^W[\RDM]}{\partial \RDM_{\alpha\beta}}
&= -h_{\beta\alpha}.
\end{align}

The constrained-search representation of Levy and Valone \cite{levy1979universal,Valone1980} expresses the finite-temperature functional as
\begin{align}
F_\beta^W[\RDM]
&=
\min_{\Gamma\rightarrow \RDM}
\Bigg[
\sum_i P_i\braket{\Psi_i|W|\Psi_i}
\nonumber\\
&\quad
+\frac{1}{\beta}\sum_i P_i\ln P_i
\Bigg],
\label{eq:constrained_search}
\end{align}
where $\Gamma=\{P_i,\Psi_i\}$ denotes an ensemble of orthonormal many-particle states $\{\Psi_i\}$ with probabilities $P_i$.  The functional is often decomposed into Hartree, exchange, noninteracting entropy, and correlation contributions \cite{Helbig2006,Bloechl2013,Baldsiefen2015},
\begin{equation}
F_\beta^W[\RDM]
=F_\mathrm{H}^W[\RDM]+F_\mathrm{x}^W[\RDM]
+F_\beta^0[\RDM]+F_{\mathrm{c},\beta}^W[\RDM].
\end{equation}
The kinetic energy is not part of $F_\beta^W$ because it is already contained in $\Tr[\RDM h]$.

\subsection{DF+RDMF decomposition}

In a DF+RDMF approach the interaction is separated into a part treated by an approximate density functional and a part treated by an explicit density-matrix functional.  A convenient starting point is obtained by adding and subtracting the same interaction functional,
\begin{equation}
F^W[\RDM]
= \FDF^W[\RDM]
+\left(
F^{W'}[\RDM]-\FDF^{W'}[\RDM]
\right),
\label{eq:df_rdmf_start}
\end{equation}
where $W'$ denotes the interaction component for which the density-matrix correction is retained.  The term in parentheses replaces the density-functional description of $W'$ by an explicit density-matrix functional.  With $\mathcal{D}_N=\{\RDM\,|\,0\leq\RDM\leq\one,\Tr[\RDM]=N\}$, the ground-state problem becomes
\begin{align}
E_N[h]
&=
\min_{\RDM\in\mathcal{D}_N}
\Bigl\{
\Tr[\RDM h]+\FDF^W[\RDM]
\nonumber\\
&\quad
+F^{W'}[\RDM]-\FDF^{W'}[\RDM]
\Bigr\}.
\label{eq:df_rdmf_energy}
\end{align}
This form makes the double-counting structure transparent: the density functional describes the full interaction approximately, and the local part $W'$ is then corrected at the RDMF level.

\section{Real-Space-Decomposed DF+RDMF}
\label{sec:realspace}

Earlier DF+RDMF formulations selected $W'$ through a subset of localized orbitals \cite{Bloechl2011}.  Such an orbital-space decomposition is intuitive for localized correlations, but it can obscure the connection between the interaction used in the density-functional term and the one used in the density-matrix correction.  We therefore define $W'$ directly by decomposing the Coulomb interaction in real space.

Let
\begin{equation}
w(\rvec,\rvec')
=\frac{e^2}{4\pi\varepsilon_0|\rvec-\rvec'|}
\end{equation}
and introduce a bounded decomposition function $0\leq \lambda(\rvec,\rvec')\leq 1$.  The interaction assigned to the density-matrix correction is
\begin{equation}
w'(\rvec,\rvec')=\lambda(\rvec,\rvec')w(\rvec,\rvec').
\label{eq:wprime}
\end{equation}
In second quantization,
\begin{equation}
W'
=\frac{1}{2}
\int \dd^4x\,\dd^4x'\,
w'(\rvec,\rvec')\,
\hat{\psi}^{\dagger}(\xvec)
\hat{\psi}^{\dagger}(\xvec')
\hat{\psi}(\xvec')
\hat{\psi}(\xvec),
\label{eq:wprime_second}
\end{equation}
where $\xvec=(\rvec,\sigma)$ includes spin.  In a one-particle basis $\{\chi_\alpha\}$ this yields
\begin{align}
W'
&=\frac{1}{2}
\sum_{\alpha\beta\gamma\delta}
U'_{\alpha\beta\gamma\delta}
\hat{c}^{\dagger}_\alpha
\hat{c}^{\dagger}_\beta
\hat{c}_\delta
\hat{c}_\gamma,\\
U'_{\alpha\beta\gamma\delta}
&=
\int \dd^4x\,\dd^4x'\,
\chi_\alpha^*(\xvec)
\chi_\beta^*(\xvec')
w'(\rvec,\rvec')
\chi_\gamma(\xvec)
\chi_\delta(\xvec').
\end{align}

The same real-space interaction is used for the density-functional contribution.  In terms of the density $n(\rvec)$ and an interaction-strength-dependent exchange-correlation hole $h_\lambda(\rvec,\rvec')$, the local interaction contribution can be written as
\begin{align}
\FDF^{W'}[n]
&=
\frac{1}{2}
\int \dd^3r\,\dd^3r'\,
n(\rvec)
\left[
n(\rvec')+h_{\lambda(\rvec,\rvec')}(\rvec,\rvec')
\right]
\nonumber\\
&\quad\times
\lambda(\rvec,\rvec')w(\rvec,\rvec').
\label{eq:local_df}
\end{align}
The essential local approximation is that the exchange-correlation hole depends on the local value of $\lambda(\rvec,\rvec')$.  This approximation keeps the density-functional and density-matrix terms tied to the same real-space interaction and avoids introducing a separate model interaction for the correction.

For spatially localized correlation effects it is useful to express $\lambda$ as a sum of local contributions,
\begin{equation}
\lambda(\rvec,\rvec')=\sum_R \lambda_R(\rvec,\rvec'),
\label{eq:lambda_sum}
\end{equation}
where each $\lambda_R$ is localized around a center $\Rvec_R$.  This gives
\begin{equation}
W'=\sum_R W_R
\end{equation}
and the DF+RDMF energy functional becomes
\begin{align}
E_N[h]
=
\min_{\substack{0\leq \RDM \leq \one\\ \Tr[\RDM]=N}}
\Bigg\{
&\Tr[\RDM h]+\FDF^W[\RDM]
\nonumber\\
&+
\sum_R
\left[
F^{W_R}[\RDM]-\FDF^{W_R}[\RDM]
\right]
\Bigg\}.
\label{eq:local_df_rdmf}
\end{align}
Possible choices include a symmetric partition
\begin{equation}
\lambda_{R}^{\mathrm{sym}}(\rvec,\rvec')
=f(|\rvec-\Rvec_R|)f(|\rvec'-\Rvec_R|)
\label{eq:lambda_sym}
\end{equation}
or a nonsymmetric partition
\begin{equation}
\lambda_{R}^{\mathrm{nonsym}}(\rvec,\rvec')
=f(|\rvec-\Rvec_R|).
\label{eq:lambda_nonsym}
\end{equation}
The specific choice should reflect the local physics targeted by the correction.

\section{Adaptive Cluster Approximation}
\label{sec:aca}

The real-space decomposition localizes the correction, but the evaluation of $F^{W_R}[\RDM]$ can still be exponentially expensive if many one-particle states remain coupled to the local interaction.  The ACA reduces this cost by rotating the bath subspace before truncation \cite{aca_schade_bloechl_2018,Schade2022ParallelQC}.  The same reduction is useful in classical calculations and in hybrid quantum-classical settings, where the retained interacting cluster determines the size of the many-particle problem, the required qubit register, and the depth of the corresponding variational state preparation.  The rotation is unitary and therefore introduces no approximation by itself; the approximation enters only when distant bath levels are discarded.

Consider a one-particle basis of dimension $N_\chi$ and assume that the local interaction acts on the first $N_{\mathrm{imp}}$ states,
\begin{equation}
W_{\mathrm{loc}}
=
\frac{1}{2}
\sum_{\alpha,\beta,\gamma,\delta\leq N_{\mathrm{imp}}}
U_{\alpha\beta\gamma\delta}
\hat{c}^{\dagger}_\alpha
\hat{c}^{\dagger}_\beta
\hat{c}_\delta
\hat{c}_\gamma.
\label{eq:wloc}
\end{equation}
The remaining $N_{\mathrm{bath}}=N_\chi-N_{\mathrm{imp}}$ states form the bath.  If the 1RDM is block diagonal,
\begin{equation}
\RDM=
\begin{pmatrix}
\RDM_{\mathrm{imp,imp}} & 0\\
0 & \RDM_{\mathrm{bath,bath}}
\end{pmatrix},
\end{equation}
then the density-matrix functional satisfies the separation property
\begin{equation}
F^{W_{\mathrm{loc}}}[\RDM]
=
F^{W_{\mathrm{loc}}}[\RDM_{\mathrm{imp,imp}}]
+F^0[\RDM_{\mathrm{bath,bath}}],
\label{eq:separation}
\end{equation}
where the bath contribution is noninteracting.  Thus the interacting many-particle problem scales with $N_{\mathrm{imp}}$ rather than with $N_\chi$.

For a general 1RDM,
\begin{equation}
\RDM=
\begin{pmatrix}
\RDM_{\mathrm{imp,imp}} & \RDM_{\mathrm{imp,bath}}\\
\RDM_{\mathrm{bath,imp}} & \RDM_{\mathrm{bath,bath}}
\end{pmatrix},
\label{eq:general_rdm}
\end{equation}
the ACA constructs a unitary transformation that acts only in the bath subspace,
\begin{equation}
U=
\begin{pmatrix}
\one & 0\\
0 & U_{\mathrm{bath,bath}}
\end{pmatrix}.
\label{eq:bath_unitary}
\end{equation}
Because the interacting subspace is left untouched, the locality of $W_{\mathrm{loc}}$ is preserved.  The transformed 1RDM,
\begin{equation}
\tilde{\rho}^{(1)}=U^\dagger \RDM U,
\end{equation}
can be brought into a block-banded form,
\begin{equation}
\tilde{\rho}^{(1)}
=
\begin{pmatrix}
\rho_{00} & \rho_{01} & 0 & \cdots\\
\rho_{10} & \rho_{11} & \rho_{12} & \cdots\\
0 & \rho_{21} & \rho_{22} & \ddots\\
\vdots & \vdots & \ddots & \ddots
\end{pmatrix},
\label{eq:banded}
\end{equation}
where block $0$ denotes the impurity and subsequent blocks represent effective bath levels.  The off-diagonal block dimensions are bounded by $N_{\mathrm{imp}}\times N_{\mathrm{imp}}$, so the correlated cluster formed by the impurity and $M$ bath levels contains at most $(M+1)N_{\mathrm{imp}}$ one-particle states.

Truncating the coupling between bath levels $M$ and $M+1$ defines the ACA($M$) hierarchy.  The local density-matrix functional is approximated as
\begin{equation}
F^{W_{\mathrm{loc}}}[\RDM]
\approx
F^{W_{\mathrm{loc}}}_{\mathrm{ACA}(M)}[\RDM]
=
F^{W_{\mathrm{loc}}}[\tilde{\rho}^{(1)}_M],
\label{eq:aca_functional}
\end{equation}
where $\tilde{\rho}^{(1)}_M$ is the block-truncated 1RDM.  The remaining exponential cost is therefore tied to the retained cluster size rather than to the full basis dimension.  In the limit of increasing $M$, the approximation systematically approaches the untruncated result.

\section{Application to Carbon Suboxide}
\label{sec:results}

Carbon suboxide has a long chemical history that makes it more than a convenient small-molecule benchmark.  It was isolated and characterized by Diels and Wolf in 1906 and soon became part of the early chemistry of ketenes and bisketenes explored by Staudinger and co-workers \cite{DielsWolf1906,StaudingerBereza1908}.  Its facile polymerization connects to the macromolecular viewpoint introduced by Staudinger, while modern materials-chemistry accounts have rediscovered C$_3$O$_2$ and its polymers as largely forgotten carbon-oxide precursors to C--H-free conjugated carbon materials and ``red carbon'' \cite{Staudinger1920,LopezSalas2020ForgottenGraphene,Odziomek2022RedCarbon}.  At the molecular level, C$_3$O$_2$ is a short oxocarbon cumulene, so its bending and electronic response are naturally related to the orbital-symmetry ideas associated with Woodward and Hoffmann and to modern analyses of helical cumulene orbitals \cite{WoodwardHoffmann1965,GarnerHoffmann2018Cumulenes}.  Most importantly for the present benchmark, far-infrared, Raman, microwave, and high-resolution rovibrational studies established C$_3$O$_2$ as a classic quasilinear or semirigid-bender molecule with a large-amplitude $\nu_7$ CCC bending coordinate \cite{Carreira1973C3O2Bending,WeberFord1976C3O2,FusinaMills1980C3O2,Bunker1980C3O2Semirigid,Lozes1981C3O2,JensenJohns1986C3O2,Masiello2005C3O2CARS}.  The literature therefore contains a genuine linear-versus-bent structural question: depending on the effective bending potential, the molecule is described as linear only after vibrational averaging, while the equilibrium potential may have a shallow bent double minimum.  Extracted barriers to linearity lie only in the range of a few tens of $\mathrm{cm^{-1}}$, approximately $20$--$80\,\mathrm{cm^{-1}}$ or $0.09$--$0.36\,\mathrm{mHa}$, i.e., roughly $0.06$--$0.2\,\mathrm{kcal\,mol^{-1}}$.  This scale is comparable to zero-point effects and much smaller than $k_{\mathrm{B}}T$ at room temperature.  Resolving this bending potential is therefore a demanding test of the electronic-structure treatment and of how accurately local correlation effects are represented.

Against this background, we consider carbon suboxide, C$_3$O$_2$, as a first molecular test.  The molecule is a compact quasilinear chain in which bending around the central carbon atom provides a sensitive probe of the local electronic structure.  The real-space-decomposed DF+RDMF formalism and the ACA reduction described above are implemented in the CP-PAW code package \cite{Bloechl2026CPPAW}.  The semilocal reference curve was obtained with the PBE generalized-gradient approximation \cite{PerdewBurkeErnzerhof1996} using the same CP-PAW framework.  In both the PBE and DF+RDMF/ACA calculations, the electronic degrees of freedom were relaxed with a damped Car--Parrinello algorithm \cite{CarParrinello1985,Kuehne2007CPMD}.  We use this implementation to assign the explicit density-matrix correction to the central carbon region, while the remaining local contributions are treated at the density-functional level.

For the symmetric real-space partition of Eq.~\eqref{eq:lambda_sym}, we choose the central carbon atom as the decomposition center and approximate the switching function by a short Gaussian expansion,
\begin{align}
f(r)
&=
e^{-1.5609r^2}
+564.35\left(e^{-1.5606r^2}-e^{-1.6503r^2}\right)
\nonumber\\
&\quad
+3729.03\left(e^{-1.6641r^2}-e^{-1.65083r^2}\right).
\label{eq:switching_function}
\end{align}
The resulting function, shown in Fig.~\ref{fig:fr}, defines a smooth local interaction region around the central carbon atom.

\begin{center}
    \centering
    \includegraphics[width=\columnwidth]{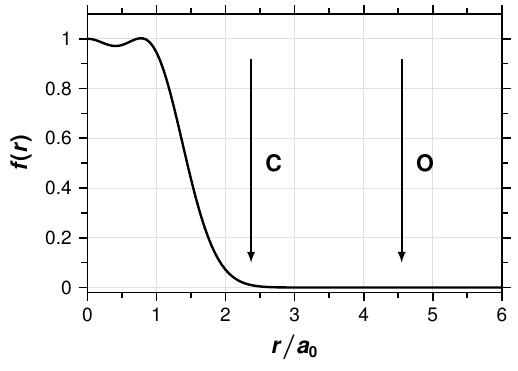}
    \captionof{figure}{Switching function $f(r)$ used in the real-space interaction decomposition.  The origin is placed at the central carbon atom of C$_3$O$_2$.  Vertical markers indicate the distances to the neighboring carbon atoms and terminal oxygen atoms.}
    \label{fig:fr}
\end{center}

\begin{center}
    \centering
    \includegraphics[width=\columnwidth]{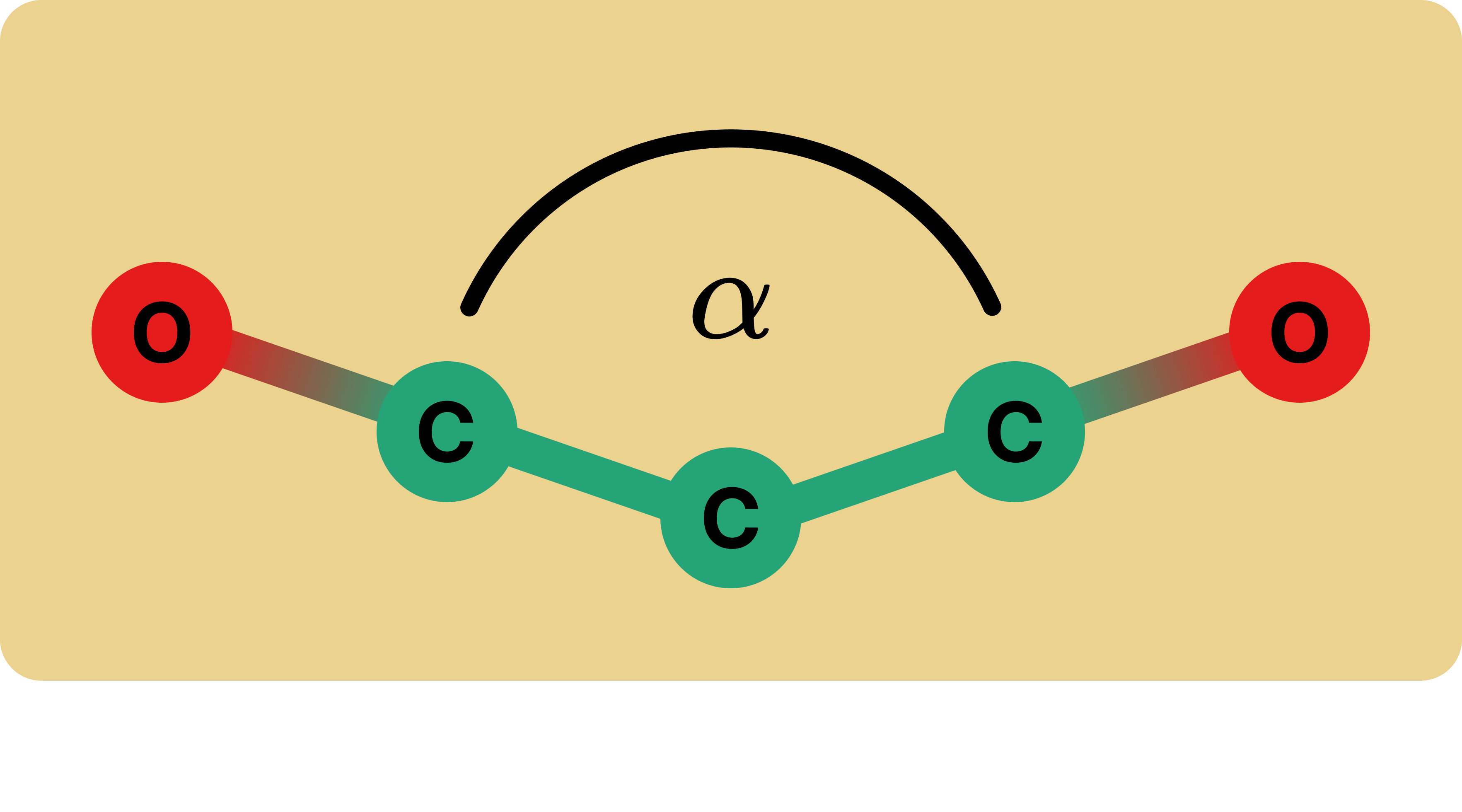}
    \captionof{figure}{Molecular structure of carbon suboxide (C$_3$O$_2$) highlighting the C--C--C bond angle $\alpha$ between the three carbon atoms.  This angle is crucial for understanding the linearity and geometry of the molecule.}
    \label{fig:structure}
\end{center}

\begin{center}
    \centering
    \includegraphics[width=\columnwidth]{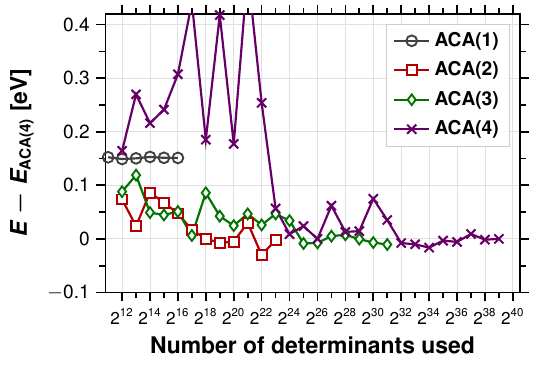}
    \captionof{figure}{Convergence of the local density-matrix correction with ACA bath level.  The data demonstrate the systematic reduction of the truncation error across the ACA(1)--ACA(4) hierarchy as additional effective bath levels are retained.}
    \label{fig:aca}
\end{center}

\begin{center}
    \centering
    \includegraphics[width=\columnwidth]{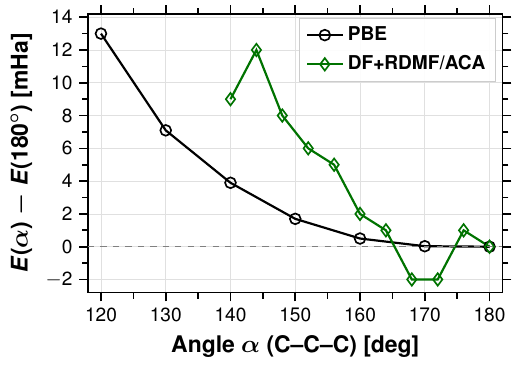}
    \captionof{figure}{Bending potentials of C$_3$O$_2$ on a common relative energy scale.  Both curves are shifted such that $E(180^\circ)=0$.  Semilocal PBE favors the linear structure, whereas the real-space-decomposed DF+RDMF/ACA selected-CI result stabilizes a bent geometry.}
    \label{fig:bending}
\end{center}

With this partition the interaction contribution is approximated as
\begin{align}
F^W[\RDM]
\approx&
\FDF^{W_{C_1}}[\RDM]
+\FDF^{W_{C_2}}[\RDM]
+\FRDMF^{W_{C_3}}[\RDM]
\nonumber\\
&+
\FDF^{W_{O_1}}[\RDM]
+\FDF^{W_{O_2}}[\RDM],
\label{eq:c3o2_partition}
\end{align}
where the explicit RDMF correction is centered on $C_3$, the middle carbon atom in the notation of the calculation.  The relevant C--C--C bending angle is defined in Fig.~\ref{fig:structure}.

The convergence of the ACA hierarchy is illustrated in Fig.~\ref{fig:aca}.  Increasing the number of retained bath levels systematically reduces the deviation of the local density-matrix correction.  For the present C$_3$O$_2$ setup, ACA(2) is already converged on the energy scale relevant for the bending potential, with residual deviations below about $0.3\,\mathrm{kcal\,mol^{-1}}$ in all conducted calculations.

Finally, Fig.~\ref{fig:bending} compares the bending potentials on a common relative energy scale by shifting each curve to $E(180^\circ)=0$.  On this scale, the semilocal PBE calculation favors the linear structure and therefore misses the shallow bent minimum implied by the quasilinear spectroscopic picture discussed above.  Thus, for this delicate bending coordinate, the mean-field-like density-functional baseline is qualitatively wrong: it predicts the wrong structural tendency rather than merely an inaccurate barrier height.

The spectroscopic literature also provides a useful angular scale for this comparison.  Because C$_3$O$_2$ is a semirigid bender with a very shallow $\nu_7$ potential, the fitted bending angles should not be interpreted as a conventional rigid equilibrium geometry; vibrational averaging can make the molecule appear nearly linear even when the effective potential has bent minima.  Nevertheless, analyses of the bending potential place the minima at sizeable but modest deviations from linearity, with reported C--C--C angles of roughly $156^\circ$--$160^\circ$, depending on the effective potential model and spectroscopic data set \cite{FusinaMills1980C3O2,Lozes1981C3O2,JensenJohns1986C3O2,Masiello2005C3O2CARS}.  The DF+RDMF/ACA curve stabilizes a bent geometry with a C--C--C angle of about $170^\circ$.  It therefore underestimates the degree of bending by about $10^\circ$--$15^\circ$ relative to these spectroscopic potentials, but it is on the correct side of the qualitative linear-versus-bent question.  The comparison is not quantitative in energy either: the calculated lowering of the bent structure is about $2\,\mathrm{mHa}$, substantially larger than the spectroscopic barrier scale of roughly $20$--$80\,\mathrm{cm^{-1}}$, or $0.09$--$0.36\,\mathrm{mHa}$.  The result should therefore be interpreted as a proof of principle that the local density-matrix correction can recover the correct bending tendency beyond semilocal DFT, while the equilibrium angle and barrier height remain imperfect in the present reduced setup.

\section{Conclusions}
\label{sec:conclusions}

We have presented a DF+RDMF framework based on a real-space decomposition of the Coulomb interaction.  In this formulation, a density functional describes the full interaction approximately, while selected local contributions are corrected by explicit reduced-density-matrix functionals.  Defining the corrected interaction directly in real space ties the density-functional and density-matrix terms to the same Coulomb partition and thereby provides a more transparent treatment of double counting than a purely orbital-based decomposition.

The remaining exponential cost of the local density-matrix functional is reduced by the adaptive cluster approximation.  Because the ACA rotates only the bath subspace, it preserves the local interaction exactly before truncation.  The approximation therefore forms a systematic hierarchy in the number of retained effective bath levels, with the correlated many-particle problem restricted to the impurity plus a small number of bath blocks.  This perspective is also aligned with recent hybrid quantum-classical RDMFT proposals, where independent local functionals can be assigned to quantum processors but still require aggressive reductions of qubit count and circuit depth \cite{Schade2022ParallelQC}.

For the C$_3$O$_2$ bending potential, semilocal PBE favors the linear structure, whereas the DF+RDMF/ACA treatment locates the minimum near a slightly bent geometry.  The failure of the mean-field-like density-functional reference is therefore qualitative, not only quantitative.  The DF+RDMF/ACA stabilization of about $2\,\mathrm{mHa}$ is still too large compared with the spectroscopic estimate of $0.09$--$0.36\,\mathrm{mHa}$, but it corrects the qualitative structural tendency that PBE gets wrong.  This result supports the central idea of the method: local density-matrix corrections can capture correlation effects beyond a density-functional description while avoiding the full exponential complexity of an uncompressed RDMF evaluation.  Future work should therefore extend the benchmark set, quantify the dependence on the real-space partition function, and test the approach for extended systems where local strong correlation and long-range density-functional screening must be combined in a size-consistent manner.

\begin{acknowledgments}
Part of the research was funded by the DFG (project numbers 417590517/CRC1415 and 519869949).
\end{acknowledgments}

\bibliography{references}

\end{document}